\documentclass[aps,prl,showpacs]{revtex4}
\usepackage{amssymb}
\usepackage{amsmath}

\usepackage{graphicx}
\usepackage{bm}

\begin{document}

\title{The Chevalley group $G_{2} (2)$ of order 12096 and the octonionic
root system of $E_{7} $}
\date{\today}
\author{Mehmet Koca}
\email{kocam@squ.edu.om}
\affiliation{Department of Physics, College of Science, Sultan Qaboos University, PO Box
36, Al-Khod 123, Muscat, Sultanate of Oman}
\author{Ramazan Ko\c{c}}
\email{koc@gantep.edu.tr}
\affiliation{Department of Physics, Faculty of Engineering University of Gaziantep, 27310
Gaziantep, Turkey}
\author{Nazife \"{O}. Koca}
\email{nazife@hotmail.com}
\affiliation{Higher College of Technology, Al-Khuwair, Muscat, Sultanate of
Oman}

\begin{abstract}
The octonionic root system of the exceptional Lie algebra $E_{8}$ has been
constructed from the quaternionic roots of $F_{4}$ using the Cayley-Dickson
doubling procedure where the roots of $E_{7}$ correspond to the imaginary
octonions. It is proven that the automorphism group of the octonionic root
system of $E_{7}$ is the adjoint Chevalley group $G_{2}(2)$ of order $12096$. One of the four maximal subgroups of $G_{2}(2)$ of order $192$ preserves
the quaternion subalgebra of the $E_{7}$ root system. The other three
maximal subgroups of orders $432$, $192$ and $336$ are the automorphism
groups of the root systems of the maximal Lie algebras $E_{6}\times U(1)$, $SU(2)\times SO(12)$ and $SU(8)$ respectively. The 7-dimensional manifolds
built with the use of these discrete groups could be of potential interest
for the compactification of the M-theory in 11-dimension.
\end{abstract}
\maketitle

\section{Introduction}

The Chevalley groups are the automorphism groups of the Lie algebras defined
over the finite fields \cite{1}. The group $G_{2}(2)$ is the automorphism
group of the Lie algebra $g_{2}$ defined over the finite field $F_{2}$ which
is one of the finite subgroups of the Lie group $G_{2}$ \cite{2}. Here we
prove that it is the automorphism group of the octonionic root system of the
exceptional Lie group $E_{7}$ .

The exceptional Lie groups are fascinating symmetries arising as groups of
invariants of many physical models suggested for fundamental interactions.
In the sequel of grand unified theories(GUT's) after $SU(5)\approx E_{4}$ %
\cite{3}, $SO(10)\approx E_{5}$ \cite{4} the exceptional group $E_{6}$ \cite%
{5} has been suggested as the largest GUT for a single family of quarks and
leptons. The 11-dimensional supergravity theory admits an invariance of the
non-compact version of $E_{7}[E_{7(-7)}]$ with a compact subgroup $SU(8)$ as
a global symmetry \cite{6}. The largest exceptional group $E_{8}$,
originally proposed as a grand unified theory \cite{7} allowing a three
family interaction of $E_{6}$, has naturally appeared in the heterotic
string theory as the $E_{8}\times E_{8}$ gauge symmetry \cite{8}.

The infinite tower of the spin representations of $SO(9)$ , the little group
of the 11-dimensional M-theory, seems to be unified in the representations
of the exceptional group $F_{4}$ \cite{9}. Moreover, it has been recently
shown that the root system of $F_{4}$ can be represented with discrete
quaternions whose automorphism group is the direct product of two binary
octahedral groups of order $48\times 48=2304$ \cite{10}.

The smallest exceptional group $G_{2}$, the automorphism group of octonion
algebra, turned out to be the best candidate as a holonomy group of the
7-dimensional manifold for the compactification of M-theory \cite{11}. For a
`` topological M-theory'' \cite{12} one may need a crystallographic
structure in 7-dimensions. In this context the root lattices of the Lie
algebras of rank-7 may play some role, such as those of $SU(8)$, $E_{7}$ and
the other root lattices of rank-7 Lie algebras. The $SU(8)$ is a maximal
subgroup of $E_{7}$ therefore it is tempting to study the $E_{7}$ root
lattice. Here a miraculous happens! The root system of $E_{7}$ can be
described by the imaginary discrete octonions \cite{13}. The Weyl group $%
W(E_{7})$ is isomorphic to a finite subgroup of $O(7)$ which is the direct
product $Z_{2}\times SO_{7}(2)$ where the latter group is the adjoint
Chevalley group of order $2^{9}.3^{4}.5.7$ \cite{14}. However, the Weyl
group $W(E_{7})$ does not preserve the octonion algebra. When one imposes
the invariance of the octonion algebra on the transformations of the $E_{7}$
roots one obtains a finite subgroup of $G_{2}$ , as expected, the adjoint
Chevalley group $G_{2}(2)$ of order $12096$ \cite{13,15}. A $G_{2}$ holonomy
group of the 7- dimensional manifold admitting the discrete symmetry $%
G_{2}(2)$ may turn out to be useful for $E_{7(-7)}$ is related to the 11-
dimensional supergravity theory.

In what follows we discuss the mathematical structure of the adjoint
Chevalley group $G_{2}(2)$ using the$126$ non-zero octonionic roots of $%
E_{7} $ without referring to its matrix representation \cite{16}.

In section 2 we construct the octonionic roots of $E_{8}$ \cite{13,17} using
the two sets of quaternionic roots of $F_{4}$ which follows the magic square
structure \cite{18} where imaginary octonions represent the roots of $E_{7}$
. First we build up a maximal subgroup of $G_{2}(2)$ of order $192$ which
preserves the quaternionic decomposition of the octonionic roots of $E_{7}$
. It is a finite subgroup of $SO(4)$ .Section 3 is devoted to a discussion
on the embeddings of the group of order $192$ in the $G_{2}(2)$. In section
4 we study the maximal subgroups of $G_{2}(2)$ and their relevance to the
root systems of the maximal Lie algebras of $E_{7}$. Finally, in section 5,
we discuss the use of our method in physical applications and elaborate the
various geometrical structures.

\section{Octonionic Root System of $E_{8}$}

In the reference \cite{13} we have shown that the octonionic root system of $%
E_{8}$ can be constructed by doubling two sets of quaternionic root system
of $F_{4}$ \cite{10} via Cayley-Dickson procedure. Symbolically we can write,%
\begin{equation}
(F_{4},F_{4})=E_{8}  \label{eq1}
\end{equation}%
where the short roots of $F_{4}$ match with the short roots of the second
set of $F_{4}$ roots and the long roots match with the zero roots. Actually (%
\ref{eq1}) follows from the magic square given by Table 1. 
\begin{table}[t]
$%
\begin{tabular}{|l|l|l|l|}
\hline
& $SU(3)$ & $SP(3)$ & $F_{4}$ \\ \hline
$SU(3)$ & $SU(3)\times SU(3)$ & $SU(6)$ & $E_{6}$ \\ \hline
$SP(3)$ & $SU(6)$ & $SO(12)$ & $E_{7}$ \\ \hline
$F_{4}$ & $E_{6}$ & $E_{7}$ & $E_{8}$ \\ \hline
\end{tabular}%
$%
\caption{Magic Square}
\label{tab:a}
\end{table}
The quaternionic scaled roots of $F_{4}$ can be given as follows:%
\begin{equation}
F_{4}:T\oplus \frac{T^{\prime }}{\sqrt{2}}  \label{eq2}
\end{equation}%
where $T\oplus T^{\prime }$ are the set of elements of the binary octahedral
group, compactly written as%
\begin{eqnarray}
T &=&{V_{0}\oplus V_{+}\oplus V_{-}}  \nonumber \\
T\prime &=&{V_{1}\oplus V_{2}\oplus V_{3}.}  \label{eq3}
\end{eqnarray}%
More explicitly, the set of quaternions $V_{0},V_{+},V_{-},V_{1},V_{2},V_{3}$
read%
\begin{eqnarray}
V_{0} &=&\left\{ {\pm 1,\pm e_{1},\pm e_{2},\pm e_{3}}\right\}  \nonumber \\
V_{+} &=&\left\{ \frac{1}{2}{\pm 1\pm e_{1}\pm e_{2}\pm e_{3}}\right\} ,%
\mathrm{even\quad number\quad of\quad (+)signs}  \label{eq4} \\
V_{-} &=&\overline{V_{+}}=\left\{ \frac{1}{2}{\pm 1\pm e_{1}\pm e_{2}\pm
e_{3}}\right\} ,\mathrm{even\quad number\quad of\quad (-)signs}  \nonumber
\end{eqnarray}%
( $\overline{V_{+}}$ is the quaternionic conjugate of $V_{+}$ )%
\begin{eqnarray}
V_{1} &=&\left\{ {\frac{1}{\sqrt{2}}(\pm 1\pm e_{1}),\frac{1}{\sqrt{2}}(\pm
e_{2}\pm e_{3})}\right\}  \nonumber \\
V_{2} &=&\left\{ {\frac{1}{\sqrt{2}}(\pm 1\pm e_{2}),\frac{1}{\sqrt{2}}(\pm
e_{3}\pm e_{1})}\right\}  \label{eq5} \\
V_{3} &=&\left\{ {\frac{1}{\sqrt{2}}(\pm 1\pm e_{3}),\frac{1}{\sqrt{2}}(\pm
e_{1}\pm e_{2})}\right\}  \nonumber
\end{eqnarray}%
where $e_{i}(i=1,2,3)$ are the imaginary quaternionic units.

Here $T$ is the set of quaternionic elements of the binary tetrahedral group
which represents the root system of $SO(8)$ and $\frac{T^{\prime }}{\sqrt{2}}
$ represents the weights of the three 8-dimensional representations of $%
SO(8) $ or, equivalently, $T$ and $\frac{T^{\prime }}{\sqrt{2}}$ represent
the long and short roots of $F_{4}$ respectively. The geometrical meaning of
these vectors are also interesting \cite{19}. Here each of the sets $%
V_{0},V_{+},V_{-}$ represent a hyperoctahedron in 4-dimensional Euclidean
space. The set $T$ is also known as a polytope $\left\{ {3,4,3}\right\} $
called 24-cell \cite{20}. Its dual polytope is $T\prime $ where $%
V_{i}(i=1,2,3)$ are the duals of the octahedron in $T$. Any two of the sets $%
V_{0},V_{+},V_{-}$ form a hypercube in 4-dimensions. Using the
Cayley-Dickson doubling procedure one can construct the octonionic roots of $%
E_{8}$ as follows:%
\begin{eqnarray}
(T,0) &=&T,(0,T)=e_{7}T  \nonumber \\
(\frac{V_{1}}{\sqrt{2}},\frac{V_{1}}{\sqrt{2}}) &=&\frac{1}{\sqrt{2}}%
(V_{1}+e_{7}V_{1})  \nonumber \\
(\frac{V_{2}}{\sqrt{2}},\frac{V_{3}}{\sqrt{2}}) &=&\frac{1}{\sqrt{2}}%
(V_{2}+e_{7}V_{3})  \label{eq6} \\
(\frac{V_{3}}{\sqrt{2}},\frac{V_{2}}{\sqrt{2}}) &=&\frac{1}{\sqrt{2}}%
(V_{3}+e_{7}V_{2})  \nonumber
\end{eqnarray}%
where $e_{1}$ , $e_{2}$ and $e_{7}$ are the basic imaginary units to
construct the other units of octonions $%
1,e_{1},e_{2},e_{3}=e_{1}e_{2},e_{4}=e_{7}e_{1},e_{5}=e_{7}e_{2},e_{6}=e_{7}e_{3} 
$ . They satisfy the algebra%
\[
e_{i}e_{j}=-\delta _{ij}+\phi _{ijk}e_{k},\quad (i,j,k=1,2,...,7) 
\]%
where $\phi _{ijk}$ is totally anti-symmetric under the interchange of the
indices $i,j,k$ and take the values $+1$ for the indices $%
123,246,435,367,651,572,741$ \cite{21}. The set of $E_{8}$ roots in (\ref%
{eq6}) can also be compactly written as the sets of octonions 
\begin{eqnarray}
\pm &&1,\frac{1}{2}(\pm 1\pm e_{a}\pm e_{b}\pm e_{c}),  \label{eq7a} \\
\pm &&e_{i}(i=1,2,...,7),\frac{1}{2}(\pm e_{d}\pm e_{f}\pm e_{g}\pm e_{h})
\label{eq7b}
\end{eqnarray}%
where the indices take $abc=123,156,147,245,267,346,357$ and $dfgh=$1246,
1257, 1345, 1367, 2356, 2347, 4567. When $\pm 1$ represent the non-zero
roots of $SU(2)$ the imaginary roots in (\ref{eq7b})which are orthogonal to $%
\pm 1$ represent the roots of $E_{7}$. The decomposition of the roots in(\ref%
{eq7a}-\ref{eq7b}) represents the branching of $E_{8}$ under its maximal
subalgebra $SU(2)\times E_{7}$ where the $112$ roots in (\ref{eq7a}) are the
weights $(\underline{2},\underline{56)}$.

A subset of roots of $F_{4}$ consisting of imaginary quaternions constitute
the roots of subalgebra $SP(3)$ with the short and long roots represented by

\begin{eqnarray}
SP(3) &:&  \nonumber \\
\mathrm{long\quad roots} &:&V_{0}^{\prime }=\left\{ {\pm e_{1},\pm e_{2},\pm
e_{3}}\right\} ;  \label{eq8} \\
\mathrm{short\quad roots} &:&\frac{V_{1}^{\prime }}{\sqrt{2}}=\left\{ \frac{1%
}{2}\left( {\pm e_{2},\pm e_{3}}\right) \right\} ,\frac{V_{2}^{\prime }}{%
\sqrt{2}}=\left\{ \frac{1}{2}\left( {\pm e_{3},\pm e_{1}}\right) \right\} ,%
\frac{V_{3}^{\prime }}{\sqrt{2}}=\left\{ \frac{1}{2}\left( {\pm e_{1},\pm
e_{2}}\right) \right\}  \nonumber
\end{eqnarray}

From the magic square one can also write the roots of $E_{7}$ in the form $%
(SP(3),F_{4})$ consisting of only imaginary octonions which can further be
put in the form%
\begin{eqnarray}
(V_{0}^{\prime },0) &=&V_{0}^{\prime },(0,T)=e_{7}T  \nonumber \\
(\frac{V_{1}^{\prime }}{\sqrt{2}},\frac{V_{1}}{\sqrt{2}}) &=&\frac{1}{\sqrt{2%
}}(V_{1}^{\prime }+e_{7}V_{1})  \nonumber \\
(\frac{V_{2}^{\prime }}{\sqrt{2}},\frac{V_{3}}{\sqrt{2}}) &=&\frac{1}{\sqrt{2%
}}(V_{2}^{\prime }+e_{7}V_{3})  \label{eq9} \\
(\frac{V_{3}^{\prime }}{\sqrt{2}},\frac{V_{2}}{\sqrt{2}}) &=&\frac{1}{\sqrt{2%
}}(V_{3}^{\prime }+e_{7}V_{2})  \nonumber
\end{eqnarray}%
The roots in (\ref{eq9}) also follows from a Coxeter-Dynkin diagram of $%
E_{8} $ where the simple roots represented by octonions depicted in Figure
1. 
\begin{figure}[h]
\begin{center}
\begin{picture}(300,220)(0,0)
\put(15,15){$\frac{1}{2}(-e_{4}+e_{5}+e_{6}-e_{7}) $}
\put(120,15){\circle*{5}}
\put(120,15){\line(0,60){30}}
\put(125,45){$-e_{6} $}
\put(120,45){\circle*{5}}
\put(120,45){\line(0,60){30}}
\put(15,75){$\frac{1}{2}(-e_{2}+e_{3}-e_{5}+e_{6}) $}
\put(120,75){\circle*{5}}
\put(120,75){\line(0,60){30}}
\put(120,75){\line(60,0){30}}
\put(155,75){$\frac{1}{2}(e_{2}-e_{3}+e_{4}-e_{7}) $}
\put(150,75){\circle*{5}}
\put(125,105){$\frac{1}{2}(e_{2}-e_{3}-e_{4}+e_{7}) $}
\put(120,105){\circle*{5}}
\put(120,105){\line(0,60){30}}
\put(15,135){$\frac{1}{2}(-e_{1}+e_{3}+e_{4}+e_{5}) $}
\put(120,135){\circle*{5}}
\put(120,135){\line(0,60){30}}
\put(125,165){$e_{1} $}
\put(120,165){\circle*{5}}
\put(120,165){\line(0,60){30}}
\put(25,195){$\frac{1}{2}(1-e_{1}-e_{2}-e_{3}) $}
\put(120,195){\circle*{5}}

\end{picture}
\end{center}
\caption{The Coxeter-Dynkin diagram of $E_{8}$ with quaternionic simple
roots }
\label{fig:one}
\end{figure}
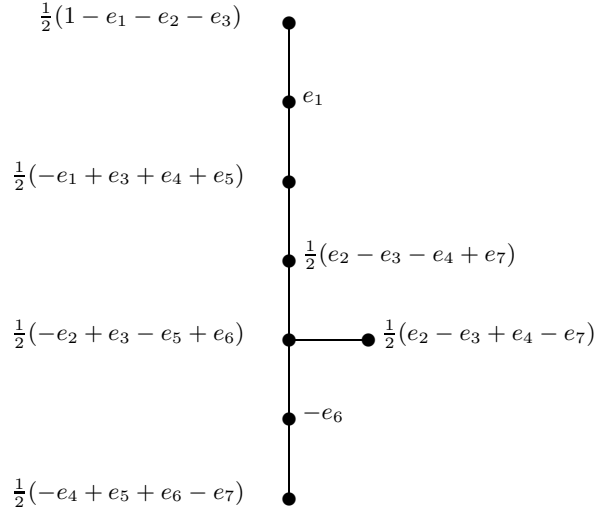
As we stated in the introduction, the automorphism group of octonionic root
system of $E_{7}$ is the adjoint Chevalley group $G_{2}(2)$, a maximal
subgroup of the Chevalley group $SO_{7}(2)$. Below we give a proof of this
assertion and show how one can construct the explicit elements of $G_{2}(2)$
without any reference to a computer calculation of the matrix representation.

We start with a theorem \cite{22} which states that the automorphism of
octonions that take the quaternions $H$ to itself form a group $[p,q]$,
isomorphic to $SO(4)\approx \frac{SU(2)\times SU(2)}{Z_{2}}$ . Here $p$ and $%
q$ are unit quaternions. In a different work \cite{23} we have studied some
finite subgroups of $O(4)$ generated by the transformations%
\begin{eqnarray}
\lbrack p,q] &:&r\rightarrow prq  \nonumber \\
\lbrack p,q]^{\ast } &:&r\rightarrow p\bar{r}q  \label{eq10}
\end{eqnarray}%
where $[p,q]$ represents an $SO(4)$ transformation preserving the norm $r%
\bar{r}=\bar{r}r$ of the quaternion $r$. More explicitly, it has been shown
in \cite{22} that the group element $[p,q]$ acts on the Cayley-Dickson
double quaternion as%
\begin{equation}
\lbrack p,q]:H+e_{7}H\rightarrow pH\bar{p}+e_{7}pHq  \label{eq11}
\end{equation}

Now we use this theorem to prove that the transformations on the root system
of $E_{7}$ in (\ref{eq9}) preserving the quaternion subalgebra form a finite
subgroup of $SO(4)$ of order $192$. In reference \cite{10} we have shown
that the maximal finite subgroup of $SO(3)$ which preserves the set of
quaternions $V\prime _{0}=\left\{ {\pm e_{1},\pm e_{2},\pm e_{3}}\right\} $
representing the long roots of $SP(3)$ as well as the vertices of an
octahedron is the octahedral group written in the form $[t,\bar{t}]\oplus
\lbrack t^{\prime },\bar{t}]$ where $t\in T$ and $t^{\prime }\in T^{\prime }$%
. On the other hand $e_{7}T$ is left invariant under the transformations $%
[p,q]\oplus \lbrack p^{\prime },q^{\prime }]$ , $(p,q\in T;p^{\prime
},q^{\prime }\in T\prime )$. Therefore the largest group preserving the
structure $(V_{0}^{\prime },0)=V_{0}^{\prime }$, $(0,T)=e_{7}T$ is a finite
subgroup of $SO(4)$ of order $576$. We will see that actually we look for a
subgroup of this group because it should also preserve the set of roots%
\begin{equation}
\frac{1}{\sqrt{2}}(V_{1}^{\prime }+e_{7}V_{1}),\frac{1}{\sqrt{2}}%
(V_{2}^{\prime }+e_{7}V_{3}),\frac{1}{\sqrt{2}}(V_{3}^{\prime }+e_{7}V_{2})
\label{eq12}
\end{equation}%
as well as keeping the form of (\ref{eq11}) invariant.

A multiplication table shown in Table 2 for the elements of the binary
octahedral group \cite{19} will be useful to follow the further discussions. 
\begin{table}[t]
$%
\begin{tabular}{|lllllll|}
\hline
& $V_{0}$ & $V_{+}$ & $V_{-}$ & $V_{1}$ & $V_{2}$ & $V_{3}$ \\ \hline
\multicolumn{1}{|l|}{$V_{0}$} & \multicolumn{1}{|l}{$V_{0}$} & $V_{+}$ & 
\multicolumn{1}{l|}{$V_{-}$} & $V_{1}$ & $V_{2}$ & $V_{3}$ \\ 
\multicolumn{1}{|l|}{$V_{+}$} & \multicolumn{1}{|l}{$V_{+}$} & $V_{-}$ & 
\multicolumn{1}{l|}{$V_{0}$} & $V_{3}$ & $V_{1}$ & $V_{2}$ \\ 
\multicolumn{1}{|l|}{$V_{-}$} & \multicolumn{1}{|l}{$V_{-}$} & $V_{0}$ & 
\multicolumn{1}{l|}{$V_{+}$} & $V_{2}$ & $V_{3}$ & $V_{1}$ \\ \cline{2-7}
\multicolumn{1}{|l|}{$V_{1}$} & \multicolumn{1}{|l}{$V_{1}$} & $V_{2}$ & 
\multicolumn{1}{l|}{$V_{3}$} & $V_{0}$ & $V_{+}$ & $V_{-}$ \\ 
\multicolumn{1}{|l|}{$V_{2}$} & \multicolumn{1}{|l}{$V_{2}$} & $V_{3}$ & 
\multicolumn{1}{l|}{$V_{1}$} & $V_{-}$ & $V_{0}$ & $V_{+}$ \\ 
\multicolumn{1}{|l|}{$V_{3}$} & \multicolumn{1}{|l}{$V_{3}$} & $V_{1}$ & 
\multicolumn{1}{l|}{$V_{2}$} & $V_{+}$ & $V_{-}$ & $V_{0}$ \\ \hline
\end{tabular}%
$%
\caption{Multiplication table of the binary octahedral group}
\label{tab:b}
\end{table}
Equation(\ref{eq11}) states that the transformation $pH\bar{p}$ fixes the
scalar part of the quaternion $H$. Therefore the transformation in (\ref%
{eq11}) acting on the root system of $E_{8}$ in (\ref{eq6}) will yield the
same result when (\ref{eq11}) acts on the roots of $E_{7}$ in (\ref{eq9}).
Now we check the transformation (\ref{eq11}) acting on the roots in (\ref%
{eq12}) and seek the form of $[p,q]$ which preserves (\ref{eq12}). More
explicitly , we look for the invariance%
\begin{eqnarray}
&&\frac{1}{\sqrt{2}}(pV_{1}^{\prime }\bar{p}+e_{7}pV_{1}q)\oplus \frac{1}{%
\sqrt{2}}(pV_{2}^{\prime }\bar{p}+e_{7}pV_{3}q)\oplus \frac{1}{\sqrt{2}}%
(pV^{\prime }\bar{p}_{3}+e_{7}pV_{2}q)  \nonumber \\
&=&\frac{1}{\sqrt{2}}(V_{1}^{\prime }+e_{7}V_{1})\oplus \frac{1}{\sqrt{2}}%
(V_{2}^{\prime }+e_{7}V_{3})\oplus \frac{1}{\sqrt{2}}(V_{3}^{\prime
}+e_{7}V_{2}).  \label{eq13}
\end{eqnarray}%
We should check all pairs in $[V_{0}\oplus V_{+}\oplus V_{-}$ , $V_{0}\oplus
V_{+}\oplus V_{-}$ ] and see that only the set of elements $[V_{0}$ , $%
V_{0}] $ , $[V_{+}$ , $V_{0}]$ , $[V_{-},V_{0}]$ satisfy the relation (\ref%
{eq13}). Just to see why $[V_{+},V_{+}]$, for example, does not work let us
apply it on the set of roots $\frac{1}{\sqrt{2}}(V_{1}^{\prime }+e_{7}V_{1})$
:%
\[
\lbrack V_{+},V_{+}]:\frac{1}{\sqrt{2}}(V_{1}^{\prime
}+e_{7}V_{1})\rightarrow \frac{1}{\sqrt{2}}(V_{+}V_{1}^{\prime
}V_{-}+e_{7}V_{+}V_{1}V_{+}). 
\]%
Using Table 2 we obtain that%
\[
\frac{1}{\sqrt{2}}(V_{1}^{\prime }+e_{7}V_{1})\rightarrow \frac{1}{\sqrt{2}}%
(V_{2}^{\prime }+e_{7}V_{1}) 
\]%
which does not belong to the set of roots of $E_{7}$. Similar considerations
eliminate all the subsets of elements in $[T,T]$ but leaves only $[T,V_{0}]$%
. Note that $[V_{+},V_{0}]^{3}=[V_{0},V_{0}]$ and it permutes the three sets
of roots of $E_{7}$ in (\ref{eq12}). Now we study the action of $[T^{\prime
},T^{\prime }]$ on the roots in (\ref{eq12}). We can easily prove that the
set of elements $[V_{1},V_{1}]$ does the job:%
\begin{eqnarray}
\lbrack V_{1},V_{1}] &:&\frac{1}{\sqrt{2}}(V_{1}^{\prime
}+e_{7}V_{1})\rightarrow \frac{1}{\sqrt{2}}(V_{1}^{\prime }+e_{7}V_{1}) 
\nonumber \\
&&\frac{1}{\sqrt{2}}(V_{2}^{\prime }+e_{7}V_{3})\leftrightarrow \frac{1}{%
\sqrt{2}}(V_{3}^{\prime }+e_{7}V_{2})  \label{eq14}
\end{eqnarray}%
We can check easily that the set of elements $[V_{2},V_{1}]$ and $%
[V_{3},V_{1}]$ also satisfy the requirements. Note that $%
[V_{i},V_{1}]^{2}=[V_{0},V_{0}]$ , $(i=1,2,3)$; any one of these set of
elements, while preserving one set of roots in(\ref{eq12}), exchange the
other two.

We conclude that the subset of elements of the group $[p,q]\oplus \lbrack
p^{\prime },q^{\prime }]$ , $(p,q\in T;p^{\prime },q^{\prime }\in T^{\prime
})$ which preserve the root system of $E_{7}$ is the group of elements $%
[T,V_{0}]\oplus \lbrack T^{\prime },V_{1}]$ of order $192$ with $17$
conjugacy classes. It is interesting to note that $[V_{0},V_{0}]$ is an
invariant subgroup of order $32$ of the group $[T,V_{0}]\oplus \lbrack
T^{\prime },V_{1}]$. Actually it is the direct product of the quaternion
group with itself consisting of elements $V_{0}=\left\{ {\pm 1,\pm e_{1},\pm
e_{2},\pm e_{3}}\right\} $. The set of elements $[T,V_{0}]\oplus \lbrack
T^{\prime },V_{1}]$ now can be written as the union of cosets of $%
[V_{0},V_{0}]$ where the coset representatives can be obtained from, say, $%
[V_{+},V_{0}]$ and $[V_{1},V_{1}]$. When $[V_{0},V_{0}]$ is taken as a unit
element then $[V_{+},V_{0}]$ and $[V_{1},V_{1}]$ generate a group isomorphic
to the symmetric group $S_{3}$. Symbolically, the group of interest can be
written as the semi-direct product of the group $[V_{0},V_{0}]$ with $S_{3}$
which is a maximal subgroup of order $576$ of the direct product of two
binary octahedral group.

It is also interesting to note that the group $[T,T]\oplus \lbrack T^{\prime
},T^{\prime }]$ has another maximal subgroup of order $192$ with $13$
conjugacy classes whose elements can be written as%
\begin{equation}
\lbrack V_{0},V_{0}]\oplus \lbrack V_{+},V_{-}]\oplus \lbrack
V_{-},V_{+}]\oplus \lbrack V_{1},V_{1}]\oplus \lbrack V_{2},V_{2}]\oplus
\lbrack V_{3},V_{3}]  \label{eq15}
\end{equation}%
This group does not preserve the root system of $E_{7}$ ,however, it
preserves the quaternion algebra in the set of imaginary octonions $\pm
e_{i}(i=1,2,...,7)$. This is also an interesting group which turns out to be
maximal in an another finite subgroup of $G_{2}(2)$ of order $1344$ \cite{24}%
. The group in (\ref{eq15}) can also be written as the semi-direct product
of $[V_{0},V_{0}]$ and $S_{3}$, however, two groups are not isomorphic
because the symmetric group $S_{3}$ here is generated by $[V_{+},V_{-}]$ and 
$[V_{1},V_{1}]$ instead of $[V_{+},V_{0}]$ and $[V_{1},V_{1}]$ as in the
previous case.

\section{63 embeddings of the quaternion preserving group in the Chevalley
group}

We go back to the equation (\ref{eq6}) and note that the binary tedrahedral
group $T=V_{0}+V_{+}+V_{-}$ played an important role in the above analysis
for it represents the root system of $SO(8)$. Any one element of the
quaternionic elements of the hypercube $V_{+}+V_{-}=\frac{1}{2}\left\{ {\pm
1\pm e_{1}\pm e_{2}\pm e_{3}}\right\} $ satisfies the relation $p^{3}=\pm 1$%
. Actually we have $112$ octonionic elements of this type in the roots of $%
E_{8}$.

We have proven in an earlier paper \cite{23}\footnote{%
It seems that this proof was given much earlier by M. Zorn \cite{26}.} that
the transformation%
\begin{equation}
b\rightarrow ab\bar{a}  \label{eq16}
\end{equation}%
where $a^{3}=\pm 1$ is an associative product of octonions which preserve
the octonion algebra. More explicitly, when $e_{i}(i=1,2,...,7)$ represent
the imaginay octonions the transformation%
\begin{equation}
e_{i}^{\prime }=ae_{i}\bar{a},(a^{3}=\pm 1)  \label{eq17}
\end{equation}%
preserves the octonion algebra%
\begin{equation}
e_{i}^{\prime }e_{j}^{\prime }=(e_{i}e_{j})^{\prime }=a(e_{i}e_{j})\bar{a}.
\label{eq18}
\end{equation}%
To work with octonionionic root systems makes life difficult because of
nonassociativity. However, the following theorem \cite{25} proves to be
useful. Let $p$ be any root of those $112$ roots and $q$ be any root of $%
E_{8}$. Consider the transformations on $q$ :%
\[
\pm p:q_{1}\equiv q,q_{2}\equiv (p)q(\bar{p}),q_{3}\equiv (\bar{p})q(p).
\]%
It was proven in \cite{25} that $q_{1},q_{2},q_{3}$ form an associative
triad $(q_{1}q_{2})q_{3}=q_{1}(q_{2}q_{3})$ satisfying the relations%
\begin{eqnarray}
&&q_{1}p\quad \mathrm{for\quad }q_{i}.\bar{p}=0,(\mathrm{42\quad triads}) 
\nonumber \\
&&q_{1}q_{2}q_{3}=\left\{ 
\begin{array}{cccc}
-1 & \mathrm{for} & q_{i}.\overline{p}=-1/2 & \mathrm{18\quad triads} \\ 
1 & \mathrm{for} & q_{i}.\overline{p}=1/2 & \mathrm{18\quad triads}%
\end{array}%
\right.   \label{eq19}
\end{eqnarray}

Actually this decomposition of $E_{8}$ roots is the same as its branching
under $SU(2)\times E_{7}$ where the non-zero roots decompose as $%
240=126+2+(2,56)$. The first $42$ triads are the $126$ non-zero roots of $%
E_{7}$ and $\pm \bar{p}$ are those of $SU(2)$. The remaining $36\times 3=108$
roots with $\pm 1,\pm p$ constitute the $112$ roots of the coset space. In
general one can show that $24$triads, out of $42$ triads, corresponding to
the roots of $E_{6}$ are imaginary octonions and the remaining $18$ triads
are those with non-zero scalar parts. The 9 triads of those octonionic roots
which satisfy the relation $q_{i}.\bar{p}=-\frac{1}{2}$ are imaginary
octonions and their negatives satisfy the relation $q_{i}.\bar{p}=\frac{1}{2}
$. When $\pm 1$ represent the roots of $SU(2)$ then all the roots of $E_{7}$
are pure imaginary as depicted in Figure 1. For a given octonion $p$ with
non-zero real part one can classify the imaginary roots of $E_{7}$ as
follows:

(i) $72$ imaginary octonions which are grouped in $24$ triads satisfying the
relation $q_{i}.\bar{p}=0$

(ii) $27$ imaginary roots classified in $9$ associative triads whose
products satisfy the relation $q_{1}q_{2}q_{3}=-1$ are the quaternionic
units. They represent the weights of the $27$ dimensional representation of $%
E_{6}$.

(iii)The remaining $9$ triads are the conjugates of those in (ii) and
represent the weights of the representation $\overline{27}$ of $E_{6}$ .

In the next section we will prove that the root system of $E_{8}$ in (\ref%
{eq6}) and equivalently those of $E_{7}$ in (\ref{eq9}) can be constructed
63 different ways preserving the octonion algebra so that the automorphism
group of the octonionic root system of $E_{7}$ is the group $G_{2}(2)$ of
order $192\times 63=12096$.

We recall that we have $18$ associative triads with non-zero scalar part,
each being orthogonal to $\bar{p}$. To distinguish the imaginary octonions
for which we keep the notation $q_{i}$ we denote the roots with non-zero
scalar part by $r_{i}$ satisfying the relation $r_{i}.\bar{p}=0$ where $%
r_{i}^{3}=\pm 1$ , $(i=1,2,3)$. They are permuted as follows :%
\[
r_{1},r_{2}=pr_{1}\bar{p},r_{3}=\bar{p}r_{1}p. 
\]%
The scalar product $r_{i}.\bar{p}=0$ can be written as%
\begin{equation}
r_{i}p+\bar{p}\bar{r}_{i}=\bar{r}_{i}p+\bar{p}r_{i}=0.  \label{eq20}
\end{equation}%
We can use (\ref{eq19}) to show that $r_{1}r_{2}=r_{2}r_{3}=r_{3}r_{1}=p$
with conjugates $\bar{r}_{2}\bar{r}_{1}=\bar{r}_{3}\bar{r}_{2}=\bar{r}_{1}%
\bar{r}_{2}=\bar{p}$.One can easily show that the octonions $r_{1}$, $r_{2}$
and $r_{3}$ are mutually orthogonal to each other:%
\begin{equation}
r_{1}.r_{2}=r_{2}.r_{3}=r_{3}.r_{1}=0\rightarrow r_{1}\bar{r}_{2}+r_{2}\bar{r%
}_{1}=r_{2}\bar{r}_{3}+r_{3}\bar{r}_{2}=r_{3}\bar{r}_{1}+r_{1}\bar{r}_{3}=0
\label{eq21}
\end{equation}%
which also implies that $r_{1}\bar{r}_{2},r_{2}\bar{r}_{3},r_{3}\bar{r}_{1}$
are imaginary octonions.

The orthogonality of $r_{1}$, $r_{2}$ and $r_{3}$ can be proven as follows.
Consider the scalar product%
\begin{equation}
r_{1}.r_{2}=\frac{1}{2}[\bar{r}_{1}(pr_{2}\bar{p})+(p\bar{r}_{1}\bar{p}%
)r_{1}].  \label{eq22}
\end{equation}%
Let us assume without loss of generality that $\bar{p}=1-p$ , $\bar{r}%
_{1}=1-r_{1}$. Substituting $\bar{p}=1-p$ and $\bar{r}_{1}=1-r_{1}$ in (\ref%
{eq22}) and using (\ref{eq20}) as well as the Moufang identities \cite{22} 
\begin{eqnarray}
(pq)(rp) &=&p(qr)p  \label{eq23a} \\
p(qrq) &=&[(pq)r]q  \label{eq23b} \\
(qrq)p &=&q[r(qp)]  \label{eq23c}
\end{eqnarray}%
one can show that $r_{1}.r_{2}=0$. Similar considerations for the other
octonions will prove that the four octonions $r_{1},r_{2},r_{3}$ and $\bar{p}
$ are mutually orthogonal to each other so that $\pm r_{1},\pm r_{2},\pm
r_{3}$ and $\pm \bar{p}$ form the vertices of a hyperoctahedron. Similarly
their conjugates forming an orthogonal quarted with their negatives
represent the vertices of another hyperoctahedron. The imaginary octonions $%
r_{1}\bar{r}_{2},r_{2}\bar{r}_{3},r_{3}\bar{r}_{1}$ are cyclically rotated
to each other in the manner $p(r_{1}\bar{r}_{2})\bar{p}=r_{2}\bar{r}_{3}$ (
cyclic permutations of $1,2,3$ ) and satisfying the relation $(r_{1}\bar{r}%
_{2}).\bar{p}=\frac{1}{2}$ where the conjugate $r_{2}\bar{r}_{1}$ satisfies
the relation $(r_{2}\bar{r}_{1}).\bar{p}=-\frac{1}{2}$. If we denote by the
imaginary octonions $E_{1}=r_{3}\bar{r}_{2},E_{2}=r_{1}\bar{r}_{3}$ and $%
E_{3}=r_{2}\bar{r}_{1}$. It is easy to prove the following identities:

\begin{eqnarray}
\bar{p} &=&\frac{1}{2}(1-E_{1}-E_{2}-E_{3})  \nonumber \\
r_{1} &=&\frac{1}{2}(1+E_{1}+E_{2}-E_{3})  \nonumber \\
r_{2} &=&\frac{1}{2}(1-E_{1}+E_{2}+E_{3})  \label{eq24} \\
r_{3} &=&\frac{1}{2}(1+E_{1}-E_{2}+E_{3})  \nonumber
\end{eqnarray}%
Therefore the set of $24$ octonions%
\begin{equation}
\left\{ \pm 1,\pm E_{1},\pm E_{2},\pm E_{3},\frac{1}{2}\left( \pm 1\pm
E_{1}\pm E_{2}\pm E_{3}\right) \right\}  \label{eq25}
\end{equation}%
are the quaternions forming the binary tedrahedral group and representing
the roots of $SO(8)$. Once this set of octonions are given we can construct
the root system of $F_{4}$ and form the roots of $E_{8}$ similar to the
equation (\ref{eq5}).

It is obvious that for a given $p(\bar{p})$ one can construct the elements
of the binary tedrahedral group, in other words, $SO(8)$ root system $9$
different ways as we have argued in the previous section. Since we have $112$
roots of this type and a choice of $p$ includes always $\pm p$ and $\pm \bar{%
p}$ that reduces such a choice to $\frac{112}{4}=28$. This number further
reduces to $\frac{28}{4}=7$ because $\bar{p},r_{1},r_{2},r_{3}$ come always
in quartets. It is not only $p(\bar{p})$ rotates $r_{1},r_{2},r_{3}$ in the
cyclic order but any one of them rotates the other three cyclically. One can
show, for example, that%
\begin{equation}
r_{1}\bar{p}\bar{r}_{1}=r_{2},r_{1}r_{2}\bar{r}_{1}=r_{3},r_{1}r_{3}\bar{r}%
_{1}=\bar{p}.  \label{eq26}
\end{equation}%
The others satisfy similar relations. Therefore the choice of elements of a
binary tedrahedral group or equivalently $F_{4}$ root system out of
octonions is $9\times 7=63$. Since the group preserving the quaternion
structure is of order $192$ the overall group which preserves the octonionic
root system of $E_{7}$ is a group of order $192\times 63=12096$. It has to
be a subgroup of $G_{2}$ and the group is certainly the Chevalley group $%
G_{2}(2)$ \cite{2}.

\section{Maximal Subgroups of $G_{2}(2)$ and the Maximal Lie Algebras of $%
E_{7}$}

There are four regular maximal Lie algebras of $E_{7} $ :

$E_{6}\times U(1)$, $SU(2)\times SO(12)$, $SU(8)$, $SU(3)\times SU(6)$; and
there are four maximal subgroups of the Chevalley group $G_{2}(2)$. It is
interesting to see whether any relations between these groups and the
octonionic root systems of these Lie algebras exist ( See M. Koca and F.
Karsch in reference \cite{2}). There is a one-to-one correspondence between
them but with one exception. When one imposes the invariance of the octonion
algebra on the root system of $SU(3)\times SU(6)$ one obtains a group which
is not maximal in the Chevalley group $G_{2}(2)$. Yet the maximal subgroup $%
[T,V_{0}]\oplus \lbrack T^{\prime },V_{1}]$ of order $192(17)$ preserves the
quaternion algebra of the magic square structure $(SP_{3},F_{4})$.The other
maximal subgroups of $G_{2}(2)$ which are of orders $432(14),192(14)$ and $%
336(9)$ have one-to-one correspondences with the groups which preserve the
octonionic root systems of $E_{6}\times U(1)$, $SU(2)\times SO(12)$ and $%
SU(8)$ respectively. In this section we will discuss the constructions of
these three maximal subgroups of $G_{2}(2)$ as the automorphism groups of
the corresponding octonionic root systems. Their character tables and the
subgroup structures can be found in reference \cite{27}.

\subsection{Octonionic root system of $E_{6}\times U(1)$ and the group of
order $432(14)$}

Since $U(1)$ factor is represented by zero root we are essentially looking
at the roots of $E_{6}$ in $E_{7}$ . Either using the simple roots of $E_{8}$
in Figure 1 or those roots of $E_{7}$ already given in equation (\ref{eq7b})
we may decompose the roots of $E_{7}$ to those roots orthogonal to the
vector $\frac{1}{2}(1-e_{1}-e_{2}-e_{3})$ which constitute the $72$ roots of 
$E_{6}$ and the ones having a scalar product $\pm \frac{1}{2}$ with it will
be the weights of the representations $\underline{27}+\underline{27^{\ast }}$%
. In an explicit form they read:

Non-zero roots of $E_{6}$ :%
\begin{eqnarray}
&&\pm e_{4},\pm e_{5},\pm e_{6},\frac{1}{2}(\pm e_{4}\pm e_{5}\pm e_{6}\pm
e_{7}),\pm \frac{1}{2}(e_{2}-e_{3}\pm e_{4}\pm e_{7}),\pm \frac{1}{2}%
(e_{2}-e_{3}\pm e_{5}\pm e_{6}),  \nonumber \\
&&\pm \frac{1}{2}(e_{3}-e_{1}\pm e_{6}\pm e_{7}),\pm \frac{1}{2}%
(e_{3}-e_{1}\pm e_{4}\pm e_{5}),  \label{eq27} \\
&&\pm \frac{1}{2}(e_{1}-e_{2}\pm e_{5}\pm e_{7}),\pm \frac{1}{2}%
(e_{1}-e_{2}\pm e_{4}\pm e_{6})  \nonumber
\end{eqnarray}%
The number in the bracket is the number of conjugacy classes and is used to
distinguish the groups having the same order.

Weights of $\underline{27}+\underline{27^{\ast }}$ of $E_{6}$ :%
\begin{eqnarray}
&&\pm e_{1},\pm e_{2},\pm e_{3}  \nonumber \\
&&\pm \frac{1}{2}(e_{2}+e_{3}\pm e_{4}\pm e_{7}),\pm \frac{1}{2}%
(e_{2}+e_{3}\pm e_{5}\pm e_{6}),  \nonumber \\
&&\pm \frac{1}{2}(e_{3}+e_{1}\pm e_{6}\pm e_{7}),\pm \frac{1}{2}%
(e_{3}+e_{1}\pm e_{4}\pm e_{5}),  \label{eq28} \\
&&\pm \frac{1}{2}(e_{1}+e_{2}\pm e_{5}\pm e_{7}),\pm \frac{1}{2}%
(e_{1}+e_{2}\pm e_{4}\pm e_{6})  \nonumber
\end{eqnarray}%
Now we are in a position to determine the subgroup of the group of order $%
192(17)$ which preserves this decomposition.

The magic square indicates that the root system of $E_{6}$ can be obtained
by Cayley-Dickson procedure as the pair $(SU(3),F_{4})$ which is clear from (%
\ref{eq27}) where the roots of $(SU3)$ are represented by the short roots $%
\pm \frac{1}{2}(e_{2}-e_{3})$ , $\pm \frac{1}{2}(e_{3}-e_{1})$ , $\pm \frac{1%
}{2}(e_{1}-e_{2})$.

It can be shown that the subgroup of the group of order $192(17)$ preserving
this system of roots where the imaginary unit $e_{7}$ is left invariant is
the group generated by the elements,%
\begin{equation}
\lbrack t,V_{0}],[\frac{1}{\sqrt{2}}(e_{2}-e_{3}),V_{1}]  \label{eq29}
\end{equation}%
Here $t$ is given by $t=\frac{1}{2}(1+e_{1}+e_{2}+e_{3})$ . More explicitly
we can write the elements of the group of interest as follows 
\begin{eqnarray}
\lbrack t,V_{0}] &\subset &[V_{+},V_{0}],[\bar{t},V_{0}]\subset \lbrack
V_{-},V_{0}],[1,V_{0}]\subset \lbrack V_{0,}V_{0}];  \label{eq30a} \\
\lbrack \frac{1}{\sqrt{2}}(e_{2}-e_{3}),V_{1}]{} &\subset &[V_{1},V_{1}],[%
\frac{1}{\sqrt{2}}(e_{3}-e_{1}),V_{1}]\subset \lbrack V_{2},V_{1}],
\label{eq30b} \\
\lbrack \frac{1}{\sqrt{2}}(e_{1}-e_{2}),V_{1}] &\subset &[V_{3},V_{1}]. 
\nonumber
\end{eqnarray}%
Each set contains $8$ elements hence the group is of order $48$. We recall
that in the decomposition of the root system of $E_{7}$ in (\ref{eq27}) and (%
\ref{eq28}) under $E_{6}$ the quaternions $\pm t(\pm \bar{t})$ and thereby
the quaternionic imaginary units $e_{1},e_{2},e_{3}$ are used. This implies
that the sum $\frac{1}{\sqrt{3}}(e_{1}+e_{2}+e_{3})$ is left invariant under
the transformations $tq\bar{t}$ where $q$ is any octonion. This proves that
the group of concern is a finite subgroup of $SU(3)$ acting in the
6-dimensional Euclidean subspace. The discussions through the relations (\ref%
{eq16}-\ref{eq19}) show that one can construct the root system of $E_{6}$ in
(\ref{eq27}), consequently those weights in (\ref{eq28}), $9$ different ways
implying that the group of order preserving the root system of $E_{6}$ in (%
\ref{eq27}) is a finite subgroup of $SU(3)$ of order $48\times 9=432$ with $%
14$ conjugacy classes. The $6\times 6$ irreducible matrix representation of
this group as well as its character table can be found in reference \cite{27}%
.

\subsection{The octonionic root system of $SU(2)\times SO(12)$ and the group
of order $192(14)$}

Existence of an automorpism group of order $192$ is obvious since the $SU(2)$
roots are any imaginary octonion $\pm q$ which must be left invariant under
any transformation. Since we have $126/2=63$ choice for the $SU(2)$ roots
the group of invariance is $12096/63=192$. The structure of this group is
totally different than the previous group of order $192(17)$ as we will
discuss below.

The magic square tells us that the root system of $SO(12)$ can be obtained
by pairing two sets of quaternionic roots of $SP(3)$ \textit{a'la}
Cayley-Dickson procedure $(SP(3),SP(3))$. When we take the quaternionic
roots of $SP(3)$ given in(\ref{eq8}) we obtain the root system of $SO(12)$
and $SU(2)$ as follows:

$SO(12)$ roots :

\begin{eqnarray}
\pm e_{1},\pm e_{2},\pm e_{3},e_{7}(\pm e_{1},\pm e_{2},\pm e_{3}) &=&\pm
e_{4},\pm e_{5},\pm e_{6}  \nonumber \\
\frac{1}{2}(\pm e_{2}\pm e_{3})+e_{7}\frac{1}{2}(\pm e_{2}\pm e_{3}) &=&%
\frac{1}{2}(\pm e_{2}\pm e_{3}\pm e_{5}\pm e_{6})  \nonumber \\
\frac{1}{2}(\pm e_{3}\pm e_{1})+e_{7}\frac{1}{2}(\pm e_{1}\pm e_{2}) &=&%
\frac{1}{2}(\pm e_{1}\pm e_{3}\pm e_{4}\pm e_{5})  \label{eq31} \\
\frac{1}{2}(\pm e_{1}\pm e_{2})+e_{7}\frac{1}{2}(\pm e_{3}\pm e_{1}) &=&%
\frac{1}{2}(\pm e_{1}\pm e_{2}\pm e_{4}\pm e_{6})  \nonumber
\end{eqnarray}

$SU(2)$ roots : $\pm e_{7}.$

The remaining roots transform as the weights of the representation $(%
\underline{2},\underline{32}^{\prime })$ under $SU(2)\times SO(12)$. Since
the root $\pm e_{7}$ remains invariant under any transformation which
preserves the decomposition of $E_{7}$ under $SU(2)\times SO(12)$ the group
which we seek is a finite subgroup of $SU(3)$. We recall from the previous
discussions that the quaternionic root system of $SP(3)$ is preserved by the
octahedral group $[T,\bar{T}]\oplus \lbrack T\prime ,T\prime ]$. However, we
seek a subgroup of $[T,V_{0}]\oplus \lbrack T\prime ,V_{1}]$ which is also a
subgroup of the octahedral group. Since we have $V_{0}$ and $V_{1}$ on the
right of the pairs it should be $[\bar{V}_{0},V_{0}]\oplus \lbrack \bar{V}%
_{1},V_{1}]$. Actually we can write all the group elements explicitly, 
\begin{eqnarray}
&&[1,1],[e_{1},-e_{1}],[\frac{1}{\sqrt{2}}(1+e_{1}),\frac{1}{\sqrt{2}}%
(1-e_{1})],[\frac{1}{\sqrt{2}}(1-e_{1}),\frac{1}{\sqrt{2}}(1+e_{1})]
\label{eq32a} \\
&&[e_{2},-e_{2}],[e_{3},-e_{3}],[\frac{1}{\sqrt{2}}(e_{2}+e_{3}),-\frac{1}{%
\sqrt{2}}(e_{2}+e_{3})],[\frac{1}{\sqrt{2}}(e_{2}-e_{3}),\frac{1}{\sqrt{2}}%
(-e_{2}+e_{3})]  \label{eq32b}
\end{eqnarray}%
The elements in (\ref{eq32b}) form a cyclic group $Z_{4}$ and those in (\ref%
{eq32b}) are the right or left cosets of (\ref{eq32b}) with, say, $%
[e_{2},-e_{2}]$ is a coset representative. Indeed the elements $[1,1]$ and $%
[e_{2},-e_{2}]$ form the group $Z_{2}$ which leaves the group $Z_{4}$
invariant under conjugation. Hence the group of order $8$ in (\ref{eq32a}-%
\ref{eq32b}) has the structure $Z_{4}:Z_{2}$ where $Z_{4}$ is an invariant
subgroup. We may also allow $e_{7}\rightarrow -e_{7}$ that amounts to
extending the group $Z_{4}:Z_{2}$ by the element $[-1,1]$. Since the element 
$[-1,1]$ commutes with the elements of $Z_{4}:Z_{2}$ then we have a group of
order $16$ with the structure $Z_{2}\times (Z_{4}:Z_{2})$. This is the group
of automorphism of the root system in (\ref{eq31}) when the quaternionic
units are taken to be $e_{1},e_{2}$ and $e_{3}$.

Now the question is how many different ways we decompose (\ref{eq31})
allowing $e_{7}\rightarrow \pm e_{7}$ only. In other words, what is the
number of quaternionic units one can choose allowing $e_{7}\rightarrow \pm
e_{7}$. These units of quaternions can be chosen from the set of $112$ roots
orthogonal to $e_{7}$. They are

\begin{equation}
\frac{1}{2}(\pm 1\pm e_{1}\pm e_{2}\pm e_{3}),\frac{1}{2}(\pm 1\pm e_{1}\pm
e_{5}\pm e_{6}),\frac{1}{2}(\pm 1\pm e_{2}\pm e_{4}\pm e_{5}),\frac{1}{2}%
(\pm 1\pm e_{3}\pm e_{4}\pm e_{6}).  \label{eq33}
\end{equation}

One can prove that each set of $16$ octonions in (\ref{eq33}) will yield to $%
3$ sets of quaternionic imaginary units not involving $e_{7}$. Therefore
there are $12$ different quaternionic units to build the group structure $%
Z_{2}x(Z_{4}:Z_{2})$ and the number of overall elements of the group
preserving the root system in (\ref{eq31}) is $12\times 16=192$. To give a
nontrivial example let us choose $p=\frac{1}{2}(1+e_{2}+e_{4}+e_{5})$ with $%
\bar{p}=\frac{1}{2}(1-e_{2}-e_{4}-e_{5})$. The following set of octonions
chosen from(\ref{eq31})%
\begin{equation}
\frac{1}{2}(\pm e_{1}+e_{2}+e_{4}\pm e_{6}),\frac{1}{2}(\pm
e_{3}+e_{2}+e_{5}\pm e_{6}),\frac{1}{2}(\pm e_{1}+e_{4}+e_{5}\pm e_{3})
\label{eq34}
\end{equation}%
have scalar products $q_{i}.\bar{p}=0$ where $q_{i}$ is one of those in (\ref%
{eq34}). Under the rotation $pq_{i}\bar{p}$, for example, the quaternionic
units%
\begin{equation}
E_{1}=\frac{1}{2}(e_{2}+e_{5}+e_{5}-e_{6}),E_{2}=\frac{1}{2}%
(e_{1}-e_{3}+e_{4}+e_{5}),E_{3}=\frac{1}{2}(-e_{1}+e_{2}+e_{4}+e_{6})
\label{eq35}
\end{equation}%
are permuted and one can construct (\ref{eq31}) with the set of octonions

$SO(12)$ roots:%
\begin{eqnarray}
\pm E_{1},\pm E_{2},\pm E_{3},e_{7}(\pm E_{1},\pm E_{2},\pm E_{3}) &=&\pm
E_{4},\pm E_{5},\pm E_{6}  \nonumber \\
\frac{1}{2}(\pm E_{2}\pm E_{3})+e_{7}\frac{1}{2}(\pm E_{2}\pm E_{3}) &=&%
\frac{1}{2}(\pm E_{2}\pm E_{3}\pm E_{5}\pm E_{6})  \nonumber \\
\frac{1}{2}(\pm E_{3}\pm E_{1})+e_{7}\frac{1}{2}(\pm E_{1}\pm E_{2}) &=&%
\frac{1}{2}(\pm E_{1}\pm E_{3}\pm E_{4}\pm E_{5})  \label{eq36} \\
\frac{1}{2}(\pm eE_{1}\pm E_{2})+e_{7}\frac{1}{2}(\pm E_{3}\pm E_{1}) &=&%
\frac{1}{2}(\pm E_{1}\pm E_{2}\pm E_{4}\pm E_{6})  \nonumber
\end{eqnarray}

$SU(2)$ roots: $\pm e_{7}$

This is certainly invariant under the quaternion preserving automorphism
group of order $16$ as discussed above where the imaginary quaternionic
units $e_{1},e_{2},e_{3}$ in (\ref{eq32a}-\ref{eq32b} ) are replaced by $%
E_{1},E_{2},E_{3}$ in (\ref{eq35}). One can proceed in the same manner and
construct $12$ different sets of quaternionic units by which one constructs
the group $Z_{2}\times (Z_{4}:Z_{2})$.

\subsection{Octonionic root system of $SU(8)$ and the automophism group of
order $336(9)$}

Using the Coxeter-Dynkin diagram of figure1 we can write the octonionic
roots of $SU(8)$ as follows:%
\begin{eqnarray}
&&\pm e_{1},\pm e_{2},\pm e_{4},\pm e_{6}  \nonumber \\
&&\frac{1}{2}(\pm e_{1}\pm e_{2}+e_{5}+e_{7}),\frac{1}{2}(\pm e_{1}\pm
e_{4}+e_{3}+e_{5}),\frac{1}{2}(\pm e_{1}\pm e_{6}+e_{3}+e_{7})  \label{eq37}
\\
&&\frac{1}{2}(\pm e_{2}\pm e_{4}+e_{3}-e_{7}),\frac{1}{2}(\pm e_{2}\pm
e_{6}-e_{3}+e_{5}),\frac{1}{2}(\pm e_{4}\pm e_{6}-e_{5}+e_{7}).  \nonumber
\end{eqnarray}%
First of all, we note that the roots of $E_{7}$ decompose under its maximal
Lie algebra $SU(8)$ as $126=56+70$. Therefore those roots of $E_{7}$ in (\ref%
{eq7b}) not displayed in (\ref{eq37}) are the weights of the $70$
dimensional representation of $SU(8)$.

To determine the automorphism group of the set in (\ref{eq37}) we may follow
the same method discussed above however here we choose a different way for $%
SU(8)$ is not in the magic square.

In an earlier paper \cite{16} we have constructed the 7-dimensional
irreducible representation of the group $PSL_{2}(7):Z_{2}$ of order $336$
and proved that this group preserves the octonionic root system of $SU(8)$.
Below we give three matrix generators of the Klein's group $PSL_{2}(7)$, a
simple group with $6$ conjugacy classes,%
\begin{eqnarray}
A &=&\frac{1}{2}\left[ 
\begin{array}{ccccccc}
-1 & -1 & 0 & 0 & -1 & 0 & -1 \\ 
0 & 0 & 0 & 1 & -1 & 1 & 1 \\ 
1 & -1 & 0 & 1 & 0 & -1 & 0 \\ 
0 & -1 & -1 & 0 & 1 & 1 & 0 \\ 
0 & -1 & 1 & -1 & 0 & 0 & 1 \\ 
1 & 0 & -1 & -1 & -1 & 0 & 0 \\ 
1 & 0 & 1 & 0 & 0 & 1 & -1%
\end{array}%
\right] ;B=\left[ 
\begin{array}{ccccccc}
-1 & 0 & 0 & 0 & 0 & 0 & 0 \\ 
0 & -1 & 0 & 0 & 0 & 0 & 0 \\ 
0 & 0 & 1 & 0 & 0 & 0 & 0 \\ 
0 & 0 & 0 & 0 & 0 & -1 & 0 \\ 
0 & 0 & 0 & 0 & 0 & 0 & 1 \\ 
0 & 0 & 0 & -1 & 0 & 0 & 0 \\ 
0 & 0 & 0 & 0 & 1 & 0 & 0%
\end{array}%
\right]  \nonumber \\
C &=&\frac{1}{2}\left[ 
\begin{array}{ccccccc}
0 & 0 & 0 & -2 & 0 & 0 & 0 \\ 
2 & 0 & 0 & 0 & 0 & 0 & 0 \\ 
0 & 0 & 0 & 0 & 0 & 0 & 2 \\ 
0 & 1 & 1 & 0 & -1 & -1 & 0 \\ 
0 & -1 & -1 & 0 & -1 & -1 & 0 \\ 
0 & -1 & 1 & 0 & -1 & 1 & 0 \\ 
0 & 1 & -1 & 0 & -1 & 1 & 0%
\end{array}%
\right]  \label{eq38}
\end{eqnarray}%
These matrices satisfy the relation%
\begin{equation}
A^{4}=B^{2}=C^{7}=I.  \label{eq39}
\end{equation}%
The matrices $A$ and $B$ generate the octahedral subgroup of order $24$ of
the Klein's group.

The $56$ octonionic roots can be decomposed into 7-sets of hyperocthedra in
4- dimensions. The matrix $C$ permutes the seven sets of octahedra to each
other. The octahedral group generated by $A$ and $B$ preserves one of the
octahedra while transforming the other sets to each other. We display the
7-octahedra as follows:%
\[
\begin{array}{ccccc}
& \pm e_{2} &  &  & \pm e_{1} \\ 
& \pm \frac{1}{2}(e_{4}-e_{5}+e_{6}+e_{7}) &  &  & \pm \frac{1}{2}%
(e_{2}+e_{3}-e_{5}+e_{6}) \\ 
\underline{1}: & \pm \frac{1}{2}(e_{1}-e_{3}+e_{6}-e_{7}) &  & \underline{2}:
& \pm \frac{1}{2}(-e_{2}+e_{3}-e_{4}-e_{7}) \\ 
& \mp \frac{1}{2}(e_{1}+e_{3}+e_{4}+e_{5}) &  &  & \pm \frac{1}{2}%
(e_{4}+e_{5}+e_{6}-e_{7})%
\end{array}%
\]%
\[
\begin{array}{ccccc}
& \mp e_{4} &  &  & \pm \frac{1}{2}(-e_{2}-e_{3}+e_{5}+e_{6}) \\ 
& \pm \frac{1}{2}(e_{1}+e_{3}+e_{6}+e_{7}) &  &  & \pm \frac{1}{2}%
(-e_{4}-e_{5}+e_{6}+e_{7}) \\ 
\underline{3}: & \pm \frac{1}{2}(-e_{1}-e_{2}+e_{5}+e_{7}) &  & \underline{4}%
: & \pm \frac{1}{2}(-e_{1}-e_{3}+e_{4}-e_{5}) \\ 
& \pm \frac{1}{2}(e_{2}-e_{3}+e_{5}+e_{6}) &  &  & \mp \frac{1}{2}%
(-e_{1}+e_{2}+e_{5}+e_{7})%
\end{array}%
\]%
\[
\begin{array}{ccccc}
& \mp \frac{1}{2}(e_{1}+e_{2}+e_{5}+e_{7}) &  &  & \pm \frac{1}{2}%
(-e_{1}+e_{3}+e_{4}+e_{5}) \\ 
& \pm e_{6} &  &  & \pm \frac{1}{2}(-e_{2}+e_{3}-e_{5}+e_{6}) \\ 
\underline{5}: & \pm \frac{1}{2}(e_{2}+e_{3}+e_{4}-e_{7}) &  & \underline{6}:
& \pm \frac{1}{2}(e_{1}+e_{3}-e_{6}+e_{7}) \\ 
& \mp \frac{1}{2}(e_{1}-e_{3}+e_{4}-e_{6}) &  &  & \mp \frac{1}{2}%
(e_{2}+e_{3}-e_{4}-e_{7})%
\end{array}%
\]%
\[
\begin{array}{cc}
& \pm \frac{1}{2}(e_{4}-e_{5}-e_{6}+e_{7}) \\ 
& \pm \frac{1}{2}(-e_{1}+e_{3}+e_{6}+e_{7}) \\ 
\underline{7}: & \pm \frac{1}{2}(e_{2}-e_{3}-e_{4}+e_{7}) \\ 
& \mp \frac{1}{2}(e_{1}-e_{2}+e_{5}+e_{7})%
\end{array}%
\]%
Note that each vector is orthogonal to the others in a given set of $8$
vectors forming an octahedron in 4-dimensions. The matrix $C$ permutes the
set of octahedra as $1\rightarrow 2\rightarrow 3\rightarrow 4\rightarrow
5\rightarrow 6\rightarrow 7\rightarrow 1.$ The matrices $A$ and $B$ leave
the set of vectors in \underline{$1$} invariant and transforms the other
sets to each others as follows:

$A:$ $2\rightarrow 5\rightarrow 6\rightarrow 7\rightarrow 2$ ; $%
3\leftrightarrow 4$ and leaves $1$ invariant.

$B:$ $3\leftrightarrow 5$ ; $4\leftrightarrow 7$ and leaves each of the set $%
1,2,6$ invariant.

When we decompose the weights of the 70 dimensional representation of $SU(8)$
under the octahedral group the vectors are partitioned as numbers of vectors 
$2,6,6,8,12,12,24$. The vector $\pm \frac{1}{2}(e_{1}+e_{2}+e_{5}-e_{7})$ is
left invariant under the octahedral group which corresponds to its 2
dimensional irreducible representation. The group $PSL_{2}(7)$ can be
further extended to the group $PSL_{2}:Z_{2}$ of order $336$ by adding a
generator which can be obtained from the transformation $e_{1}\rightarrow
-e_{1},e_{2}\rightarrow e_{2},e_{4}\rightarrow e_{4}$ . One can readily
check that the this transformation leaves the root system of $SU(8)$
invariant.

\section{Conclusion}

We have constructed the root system of $E_{8}$ from the quaternionic roots
of $F_{4}$ \textit{a'la} Cayley-Dickson doubling procedure that is a
different realization of the magic square. The roots of $E_{7}$ are
represented by the imaginary octonions which can be constructed by doubling
the quaternionic roots of $SP(3)$ and $F_{4}$. The Weyl group of $E_{7}$ is
isomorphic to the finite group $Z_{2}\times SO_{7}(2)$ where $SO_{7}(2)$ is
the adjoint Chevalley group over the finite field $F_{2}$. We have proven
that the automorphism group of the octonionic root system of $E_{7}$ is the
adjoint Chevalley group $G_{2}(2)$, a finite subgroup of the Lie group $%
G_{2} $ of order $12096$. First we have determined one of its maximal
subgroup of order 192 which preserves the quarternion subalgebra in the root
system of $E_{7}$ and proven that this group can be embedded in the larger
group $63$ different ways. The other three maximal subgroups of orders $%
432,192,336$ respectively corresponding to the automorphism groups of the
octonionic root systems of the maximal Lie algebras $E_{6}$, $SU(2)\times
SO(12)$, $SU(8)$ have been studied in some depth. The root system of $SU(8)$
has a fascinating geometrical structure where the roots can be decomposed as
7 hyperoctahedra in 4-dimensions which are permuted to each other by one of
the generators of the Klein's group $PSL_{2}(7)$.

Any one of these subgroups or the whole group $G_{2}(2)$ could be used to
construct the manifolds which can be useful for the compactification of a
theory in $11$ dimension.

\end{document}